\documentclass[aps, prc, amsmath, amssymb, nofootinbib, twocolumn, superscriptaddress]{revtex4-1}

\usepackage[utf8]{inputenc}
\usepackage{CJK}
\usepackage[usenames]{color}
\usepackage{graphicx}
\usepackage{amsmath}
\usepackage{amsfonts}
\usepackage{amssymb}
\usepackage{mathrsfs}
\usepackage{bm}
\usepackage{verbatim}
\usepackage{cancel}
\usepackage{comment}
\usepackage[normalem]{ulem}
\usepackage[dvipsnames]{xcolor}
\usepackage{diagbox}
\usepackage{slashed}
\usepackage{subfigure}

\usepackage{hyperref}
\hypersetup{
  colorlinks=true,        % false: boxed links; true: colored links
  linkcolor=ForestGreen,    % color of internal links
  citecolor=PineGreen,    % color of links to bibliography
}

\newcommand{\mhi}{M_{\text{hi}}}

\newcommand{\chn}[3]{{{}^{#1}\!{#2}_{#3}}}
\newcommand{\cs}[2]{\chn{#1}{S}{#2}}
\newcommand{\cp}[2]{\chn{#1}{P}{#2}}
\newcommand{\cd}[2]{\chn{#1}{D}{#2}}

\newcommand{\csd}{{\cs{3}{1}-\cd{3}{1}}}

\begin{document}

\begin{CJK*}{UTF8}{}
\CJKfamily{gbsn}

\title{Systematic study of large-momentum distribution in nuclei with the operator product expansion}

\author{Jiexin Yu (余杰鑫)}
\email{jiexin\_yu@stu.pku.edu.cn}
\affiliation{School of Physics, Peking University, Beijing 100871, China}

\author{Bingwei Long (龙炳蔚)}
\email{bingwei@scu.edu.cn}
\affiliation{College of Physics, Sichuan University, Chengdu 610065, Sichuan Province, China}
\affiliation{Southern Center for Nuclear-Science Theory (SCNT), Institute of Modern Physics, Chinese Academy of Sciences, Huizhou 516000, Guangdong Province, China}

\date{June 5, 2025}

\begin{abstract}
The operator product expansion (OPE) is applied in conjunction with Pionless effective field theory to study the short-range structure of nuclei. 
By matching the OPE with the selected nuclear potentials for nucleon-nucleon scattering states, we obtain the Wilson coefficients. 
The nucleon momentum distribution in the deuteron is then used to test the OPE against the predictions of these nuclear potentials. 
In order to achieve a systematic separation of short-range and long-range interactions, we discuss how the OPE approximation can be improved by including higher-order EFT potentials and higher-dimension local operators. 
\end{abstract}

\maketitle

\end{CJK*}

\section{Introduction\label{sec:intro}}

The structure of atomic nuclei is dominated by interactions at a relatively long distance--- the pion Compton wavelength $m_\pi^{-1} \simeq 1.4$ fm--- between two or three nucleons. 
This is evident in ubiquitous applications of chiral effective field theory (ChEFT) in which the one-pion exchange potential is the leading-order (LO) force~\cite{Ordonez:1995rz, vanKolck:1994yi, Epelbaum:2002vt, Entem:2003ft, Navratil:2007we, Hagen:2010gd, Roth:2011ar, Ekstrom:2015rta}. 
For light nuclei, Pionless EFT in which the pions are ``integrated out'' seems to grasp their structure well~\cite{Bedaque:1999ve, Platter:2004zs, Konig:2016utl, Lu:2018bat, Konig:2019xxk, Deltuva:2020aws}. 
This indicates an even longer characteristic scale of the nuclear forces: the $S$-wave scattering lengths $\simeq 5$ or $-20$ fm. 
Phenomenological nuclear potentials that describe nucleon-nucleon scattering data toward quite high energies go through a softening procedure before entering many-body solvers; this procedure often applies a unitary transformation to decouple low- and high-resolution modes~\cite{Jurgenson:2010wy, Furnstahl:2013oba}. 
In summary, soft nuclear interactions have become favored in \textit{ab initio} calculations of nuclear structure. 
However, this creates a dilemma for studies in which external probes of the nuclei are short-ranged in nature. 
One will have to juggle the soft nuclear Hamiltonian that concerns the long-range motions with the hard probe that focuses on the degrees of freedom at short distances. 
We attempt to address this mismatch by combining the operator product expansion (OPE) with EFTs. 
The quantity studied in the present paper is the single-nucleon momentum distribution in nuclei. 
In particular, we wish to use the OPE to calculate the large-momentum part of the distribution that is attributed to nucleon-nucleon ($NN$) short-range correlations (SRCs).

For a better understanding of the short-range structure of nuclei, SRCs have been studied over the past few decades~\cite{CiofidegliAtti:2015lcu, Hen:2016kwk}. 
Short-range correlations refer to short-distance clustering of a few nucleons in nuclei, usually investigated experimentally by medium-energy proton scattering~\cite{Tang:2002ww, Piasetzky:2006ai} or electron scattering~\cite{CLAS:2003eih, CLAS:2005ola, JeffersonLabHallA:2007lly, Fomin:2011ng, Hen:2014nza, LabHallA:2014wqo}. 
From these experiments, there is already abundant evidence for the existence of two-nucleon SRCs, while the three-nucleon SRCs are under investigation~\cite{HallA:2017ivm, 3Nexp23, 3NexpLi23}.

There exist several theoretical approaches to SRCs in nuclei. 
In Refs.~\cite{Tropiano:2021qgf, Tropiano:2024bmu}, the author propose to apply the same unitary transformation that softens the microscopic nucleon-nucleon potential to the short-range operators under consideration so that the transformed operators are compatible with low-resolution wave functions.
However, it is not immediately clear how information of lattice quantum chromodynamics (QCD) can be incorporated in this approach, if it becomes available. 
Another approach is the generalized contact formalism (GCF) in which many-body wave functions are assumed to factorize into products of a short-range $NN$ part and a long-range part~\cite{Weiss:2014gua, Weiss:2015mba, Weiss:2016obx, Weiss:2023kgy}.

One can also try to calculate nuclear observables with a hard probe by solving the Schr\"odinger equation directly without applying any unitarity transformation to the nuclear potentials. 
The variational Monte Carlo (VMC) method has been used to calculate nucleon momentum distributions in various light nuclei with the Argonne V18 (AV18) potential~\cite{AV18UXVMC14} and a ChEFT potential~\cite{ChiralVMC24}. 
It will be quite challenging to perform the same calculations for heavy nuclei. 
In addition, lack of short-distance degrees of freedom other than the nucleons, such as heavy mesons or quarks and gluons, introduces model dependence, even though these nuclear potentials describe nucleon-nucleon scattering data up to the laboratory energy $T_{\rm{lab}} = 350$ MeV~\cite{Wiringa:1994wb}. 
These results for light nuclei nonetheless provide valuable benchmarks. 
One can view, say, the AV18 potential as the ``true'' underlying theory of nuclear forces, and test other theoretical approaches to SRCs against its prediction.

We will study how the OPE can be integrated into the EFTs to describe the single-nucleon momentum distribution to which SRC physics is closely related. 
The conceptual foundation of the GCF is actually similar to that of the OPE. 
They both assume some sort of factorization of long and short-range components of correlation functions, based on the idea that the short-range few-body dynamics is independent of long-range structure of the states. 
In the extreme case of unitary Fermi gases~\cite{Braaten:2008uh, Weiss:2014gua, Emmons16}, the two formalisms produce essentially the same results, both reproducing the large-momentum tail that falls off as $1/k^4$ proposed in Ref.~\cite{Tan:2008ypg}. 
Although unitarity limit is an important feature of nuclear physics and remains hopeful as a baseline for building nuclear interactions~\cite{Konig:2016utl, Konig:2019xxk}, there are other factors to be accounted for, such as range corrections and pion-exchange forces. 
When equipped with the EFT machinery the OPE can be versatile and systematic, suitable to treat these effects. 
For instance, one can choose to combine the OPE with Pionless EFT or ChEFT. 
With Pionless EFT, one takes advantage of analytic expressions readily available for the $NN$ interactions. 
When combined with ChEFT, the OPE method can be applied to more tightly bound nuclei that may not be accessible to Pionless EFT. 
Another benefit is that the pion degrees of freedom can be incorporated systematically in a manner consistent with spontaneously broken chiral symmetry of QCD~\cite{Chen:2016bde}. 
Other works utilizing the OPE to study short-distance dynamics of nonrelativistic fermions can be found in Ref.~\cite{Huang:2018yyf}.

In this paper, we will use the OPE in conjunction with Pionless EFT to calculate the large-momentum component of the single-nucleon momentum distribution in the deuteron. 
In particular, we explain how to determine the Wilson coefficients of local Pionless EFT operators that appear in the OPE. 
Making use of insensitivity of the Wilson coefficients to the external states, we choose $NN$ scattering states for matching the OPE-Pionless EFT onto the underlying nuclear potential. 
This enables us to access multiple Wilson coefficients by choosing various kinematic points for the scattering states. 
When inputs from lattice QCD calculations for similar hard probes of the $NN$ system become available, adaption of our technique will be straightforward. 

The paper is organized as follows. 
In Sec.~\ref{sec:momdis} we introduce the single-nucleon momentum distribution, followed by the OPE-Pionless EFT framework in Sec.~\ref{sec:ope} and Sec.~\ref{sec:PLEFT}. 
Comparison with various $NN$ potential models is carried out in Sec.~\ref{sec:models}.
Finally, discussions and a conclusion are offered in Sec.~\ref{sec:con}.

\section{Single-nucleon momentum distribution\label{sec:momdis}}

Nucleon momentum distributions in a nucleus provide useful insights into the multinucleon bound system. 
We write relevant quantities in terms of the nonrelativistic nucleon field, an isospinor comprised of the proton and neutron fields, each a two-component spinor itself:
\begin{equation}
    N=\left(\begin{array}{c}
        p\\
        n
    \end{array}
    \right)\, .
\end{equation}
The operator defining the single-nucleon momentum distribution is given by the Fourier transform of the nonlocal product of a distanced pair of nucleon fields:
\begin{equation}
\Omega(\vec{k}) \equiv \int d^3re^{-i\vec{k}\cdot\vec{r}} N^{\dagger}\left(-\frac{1}{2}\vec{r}\right)N\left(+\frac{1}{2}\vec{r}\right) \, .\label{eqn:omegak}
\end{equation}
For nucleus $A$ at rest with total spin $J$, isospin $T$, isospin projection $M_T$, the single-nucleon momentum distribution $\rho(k)$ is the expectation value of $\Omega(\vec{k})$ in the ground state averaged over spin projection $M_J$:
\begin{align}
        &\rho(k) \equiv \frac{1}{2J+1} \nonumber \\
        & \quad \quad \times \sum_{M_J}
         \langle A;J M_J, TM_T |\Omega(\vec{k})|A;J M_J, TM_T\rangle 
        \, ,\label{eqn:rhok}   
\end{align}
where $\rho$ depends only on the magnitude of $\vec{k}$ but not on its direction. We show in more detail in Appendix~\ref{app:RotOmega} why this is the case.

\begin{figure}
  \centering
\subfigure[]{
\includegraphics[scale=0.5]{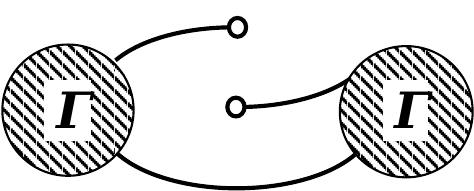}
\label{fig:nonlocalDL}
}
\subfigure[]{
\includegraphics[scale=0.5]{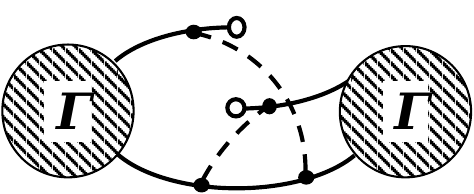}
}
\caption{
Feynman-diagram representation of the deuteron single-nucleon momentum distribution. The solid lines are the nucleon propagators and the dashed lines the meson propagators.
}
\label{fig:FeynRhok}
\end{figure}

Let us turn to the focus of the paper, the single-nucleon momentum distribution in the deuteron. 
A hadronic model $\rho(k)$ of the deuteron can be represented by the Feynman diagram Fig.~\ref{fig:FeynRhok}(a), where $\Gamma$ is the three-point vertex function~\cite{Kaplan99}
and can be related to the nuclear wave function of the deuteron. 
The dynamical propagation of mesons or other hadronic degrees of freedom, as demonstrated by Fig.~\ref{fig:FeynRhok}(b), should in principle be accounted for at some level. 
However, since Pionless EFT is adopted in the paper, in which all the mesons are integrated out, the contributions from Fig.~\ref{fig:FeynRhok}(b) do not need to be considered explicitly. Instead, those effects are either buried in the low-energy constants of the EFT or inherited from the underlying theory through the Wilson coefficients.

$\Gamma$ is obtained by solving the homogeneous Lippmann-Schwinger (LS) equation, which has the following partial-wave projected form:
\begin{equation}
      \Gamma_l(k;-B_d)=\sum_{l^{\prime}}\frac{1}{2\pi^2}\int dq q^2\frac{V_{ll^{\prime}}(k,q)}{-B_d-\frac{q^2}{m}}\Gamma_{l^{\prime}}(q;-B_d)
    \, ,\label{eqn:Gammaequ}
\end{equation}
where $B_d$ is the deuteron binding energy, $m$ the nucleon mass, $l$ the orbital angular momentum ($0$ or $2$), $q$ and $k$ are the relative momenta, and $V_{ll^{\prime}}$ is the partial-wave decomposition of the $NN$ potential. $\Gamma_{l}(k;-B_d)$ differs from the momentum-space wave function only by a noninteracting $NN$ propagator:
\begin{equation}
      \Psi_l(k;-B_d) = \frac{\Gamma_l(k; -B_d)}{-B_d-\frac{k^2}{m}} \, .
\end{equation}
The diagrammatic representation of $\rho(k)$ in Fig.~\ref{fig:FeynRhok}(a) translates into 
\begin{equation}
        \rho(k) = \sum_l \frac{\Gamma_l(k; -B_d) \Gamma_l(k; -B_d)}{(-B_d-\frac{k^2}{m})^2}\, ,
    \label{eqn:rhokGam}
\end{equation}
which we choose to normalize by the following identity:
\begin{equation}
        \int dk k^2\rho(k)=1\, .\label{eqn:NormRhok}
\end{equation}
In Sec.~\ref{sec:realistic} we will use various nuclear potentials as the underlying theory for strong interactions to calculate the momentum distribution $\rho(k)$. Since they do not have any other degrees of freedom than the nucleons, it is appropriate to use Eq.~\eqref{eqn:rhokGam}. 

In experiments, the single-nucleon momentum distributions of the deuteron can be extracted from the quasielastic inclusive electron-deuteron scattering cross-sections~\cite{Fomin:2011ng}. 
We define
\begin{equation}
    q^{\mu}=(\nu, \vec{q})\, ,\;Q^2=-q^{\mu}q_{\mu}\,,
\end{equation}
where $q^{\mu}$ is the four-momentum of the virtual photon.
For high-$Q^2$ quasi-elastic scattering with no final-state interactions, a scaling function can be obtained from the inclusive cross sections:
\begin{equation}
F(y, Q^2)=\frac{d^{2}\sigma}{d\Omega d\nu}(\sigma_{p}+\sigma_{n})^{-1}\frac{q}{[m^{2}+(y + q)^{2}]^{\frac{1}{2}}}\, ,
\end{equation}
where $\sigma_{p(n)}$ is the electron-nucleon elastic cross sections, $\Omega$ is the solid angle, and $q=|\vec{q}|$. This formula was derived with the plane-wave impulse approximation~\cite{Day:1990mf}. For sufficiently large $Q^2$, the scaling function depends only on $y$~\cite{Day:1987az, CiofidegliAtti:1990rw}, and is related to the nucleon momentum distribution of the deuteron by
\begin{equation}
    \frac{dF(k)}{dk}\approx-2\pi k\rho(k)\, .
\end{equation}

\section{operator product expansion and matching\label{sec:ope}}

We are interested in the large-momentum part of $\rho(k)$, $k \gg k_F$ where $k_F$ is the Fermi momentum of the nucleus.  
The OPE formalism elucidates separation between $k$ and $k_F$ by expanding nonlocal operators like $N^{\dagger}\left(-\frac{1}{2}\vec{r}\right)N\left(+\frac{1}{2}\vec{r}\right)$ into a series of composite local operators with state-independent coefficients:
\begin{equation}
    N^{\dagger}\left(-\frac{1}{2}\vec{r}\right)N\left(+\frac{1}{2}\vec{r}\right)
=\sum_n {W}_n(r) \mathcal{O}_n(0)
 \;\; \text{for}\; r \to 0 \, ,
\label{eqn:OPEScheme}
\end{equation}
where local EFT operators $\mathcal{O}_n(0)$ include the likes of $N^\dagger N(0)$, $(N^\dagger N)^2(0)$ and so on, $(0)$ refers to $\vec{r} = 0$, and ${W}_n(r)$ are the Wilson coefficients. This series can be organized according to the size of each term; ${W}_n(r)$ are typically in powers of $r$ multiplied by other smooth functions of $r$. $\mathcal{O}_n$ has varied mass dimension, correspondingly.

By construction the local operators $\mathcal{O}_n$ belong to an EFT that is suitable for describing long-range structure of the nucleus. Pionless EFT is our choice in the present paper. 
It describes light nuclei surprisingly well, 
and is amenable to analytic calculation in the two-body sector. 
In Pionless EFT, observables are expanded in $Q/\mhi$ where $Q$ refers generically to the external momenta in the processes of interest or infrared parameters such as the inverse $NN$ scattering lengths $a^{-1}$ in $\cs{1}{0}$ and $\cs{3}{1}$, and $\mhi \sim m_\pi$ is the breakdown scale of Pionless EFT. 

The most useful feature of the OPE in applications to SRC physics is that ${W}_n(r)$ are independent of the states between which the matrix elements of $N^{\dagger}\left(-\frac{1}{2}\vec{r}\right)N\left(+\frac{1}{2}\vec{r}\right)$ are evaluated. 
If one can somehow obtain ${W}_n(r)$, $\rho(k)$ can then be expanded as the following:
\begin{align}
  \rho(k) &= \frac{1}{2J+1} \sum_{n, M_J} \widetilde{W}_n(k) \langle A |\mathcal{O}_n^{(\slashed{\pi})}(0)|A \rangle_{M_J}  
  \, , \label{eqn:OPERhok1}
\end{align}
where $\mathcal{O}_n^{(\slashed{\pi})}$ are the Pionless local operators, the full list of quantum numbers is abbreviated to $M_J$, and the Fourier transform of the Wilson coefficients is given by
\begin{equation}
    \widetilde{W}_n(k) \equiv \int d^3re^{-i\vec{k}\cdot\vec{r}} W_n(r)\, .
\end{equation}
The OPE of $\rho(k)$ \eqref{eqn:OPERhok1} is a double expansion. 
With increasing mass dimension of $\mathcal{O}_n^{(\slashed{\pi})}$, $\widetilde{W}_n(k)$ is usually more suppressed by powers of $Q/k$, and the low-energy quantity $\langle A |\mathcal{O}_n^{(\slashed{\pi})}|A \rangle$, computable with Pionless EFT, is expanded in $Q/\mhi$. 
When $\langle A |\mathcal{O}_n^{(\slashed{\pi})}|A \rangle$ is corrected at each subleading order in the EFT, its Wilson coefficient $\widetilde{W}_n(k)$ gains correction accordingly. Therefore, the OPE/EFT representation of $\rho(k)$ can be more generally written as
\begin{equation}
    \begin{split}
  \rho(k) \propto &\sum_{n} \widetilde{W}_n^\text{LO}(k) \langle A |\mathcal{O}_n^{(\slashed{\pi})}|A \rangle_{\text{LO}} \\
  &\quad + \widetilde{W}_n^\text{LO}(k) \langle A |\mathcal{O}_n^{(\slashed{\pi})}|A \rangle_{\text{NLO}} \\
  &\quad + \widetilde{W}_n^\text{NLO}(k) \langle A |\mathcal{O}_n^{(\slashed{\pi})}|A \rangle_{\text{LO}} + \cdots
  \, , \label{eqn:OPERhok2}        
    \end{split}
\end{equation}
where $\langle A |\mathcal{O}_n^{(\slashed{\pi})}|A \rangle_{\nu\text{LO}}$ indicates that the matrix element is perturbatively calculated at the $\nu$-th order in Pionless EFT. 
The selection of the operator set $\mathcal{O}_n^{(\slashed{\pi})}$ is not unique; a linear transformation can turn them into another perfectly acceptable set of operators, and their Wilson coefficients will change correspondingly.

$\widetilde{W}_n(k)$ is determined by matching the OPE/EFT with the ``underlying'' nuclear theory.
We will evaluate the matrix elements of $\Omega(\vec{k})$ between $NN$ stationary scattering states both in the underlying nuclear theory and in the OPE/EFT. With the underlying theory, the matrix element of $\Omega(\vec{k})$ is exact, denoted by
\begin{align}
    \mathcal{A}_{\alpha' \alpha}(k; {{p}} j )\equiv  \langle \psi^{-}_{{{p}} j \alpha'}|\Omega(\vec{k})|\psi^{+}_{{{p}} j \alpha}\rangle\,, 
    \label{eqn:nonlocalmtxele}
\end{align}
where $\psi_{{{p}} j \alpha}^+$ ($\psi_{{{p}} j \alpha'}^-$) is the in (out) state with total angular momentum $j$ and the center-of-mass momentum ${{p}}$, and $\alpha$ refers collectively to other quantum numbers such as the total isospin $t$ and its projection $M_t$. 

One needs to make a tactical choice of $(p j \alpha)$ so that the desired Wilson coefficients can be extracted. 
For the moment, it suffices to state that since the deuteron is what we are interested in, we can choose $\psi^\pm_{j \alpha}$ that have also spin 1 and isospin 0. For $NN$ scattering, this is the coupled channel of $\csd$. 
It is useful to relate 
$\psi^{+}_{{{p}} j\alpha}$ 
to the off-shell $T$-matrix:
\begin{equation}
\begin{split}
    &|\psi^{+}_p; j l\rangle=|{{p}}; j l \rangle\\
    &\quad+\sum_{l^{\prime}}\frac{1}{2\pi^2}\int dqq^2\frac{T_{ll^{\prime}}({{p}},q;E)}{E-\frac{q^2}{m}+ i\epsilon}|q; j l^{\prime}\rangle\, ,    
\end{split}
\end{equation}
where the center-of-mass energy 
$E = p^2/m$ and $|{{p}}; j l\rangle$ is the noninteracting state. 
$T_{l^{\prime}l}({{p}}, q ;E)$ is obtained by solving 
the coupled-channel inhomogeneous LS equation:
\begin{equation}
\begin{split}
&T_{l^{\prime}l}(p^{\prime}, p ;E)=V_{l^{\prime}l}(p^{\prime}, p)\\
&\quad +\sum_{l^{\prime\prime}}\frac{1}{2\pi^2}\int dq q^2\frac{V_{l^{\prime}l^{\prime\prime}}(p^{\prime}, q)}{E-\frac{q^2}{m}+i\epsilon}T_{l^{\prime\prime}l}(q, p ;E)\, .
\label{eqn:InHomoLS}
\end{split}
\end{equation} 

\begin{figure}
    \centering
      \subfigure[]{
\includegraphics[scale=0.4]{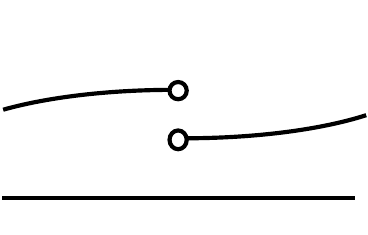}
\label{fig:nonlocal1}
}
\subfigure[]{
\includegraphics[scale=0.4]{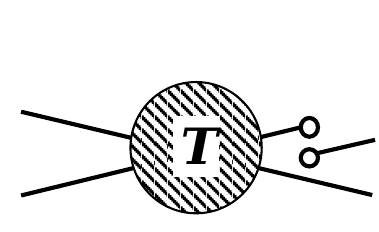}
\label{fig:nonlocal2}
}
\subfigure[]{
\includegraphics[scale=0.4]{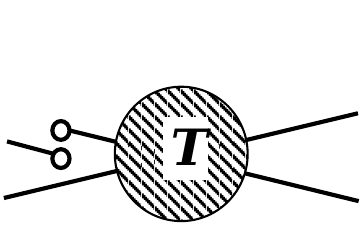}
\label{fig:nonlocal3}
}
\subfigure[]{
\includegraphics[scale=0.4]{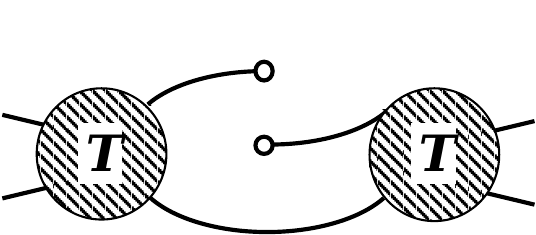}
\label{fig:nonlocal4}
}
    \caption{
    Diagrams for the matrix elements of the nonlocal operator between scattering states.}
    \label{fig:nonlocalope}
\end{figure}

With a nuclear potential $V$, 
$\mathcal{A}_{\alpha' \alpha}(k; {{p}} j)$ is represented by the Feynman diagrams in Fig.~\ref{fig:nonlocalope}. 
The contribution from the disconnected diagrams Fig.~\ref{fig:nonlocalope}(a-c) vanishes for $k \gg p$:
\begin{equation}
    \delta^{(3)}(k-{{p}})\left[(2\pi)^3\delta^{(3)}(0)+\frac{T_{l^{\prime}l}({{p}}, k ;E)+T_{l^{\prime}l}(k, {{p}} ;E)}{E-\frac{k^2}{m}}\right] \, .\label{eqn:Anonlocal_trivial}
\end{equation}
The nontrivial contribution comes from Fig.~\ref{fig:nonlocalope}(d):
\begin{equation}
    \mathcal{A}_{l^{\prime}l}(k,{{p}})=\sum_{l^{\prime\prime}}\frac{T_{l^{\prime}l^{\prime\prime}}({{p}}, k ;E)T_{l^{\prime\prime}l}(k, {{p}} ;E)}{(E-\frac{k^2}{m})^2}\, .
    \label{eqn:ANonLocal}
\end{equation}

On the other hand, $\mathcal{A}_{\alpha' \alpha}(k; {{p}} j )$ can be calculated with the OPE and Pionless EFT: 
\begin{equation}
\mathcal{A}_{\alpha' \alpha}(k; {{p}} j) =
\sum_n \widetilde{W}_{n}(k)
\mathcal{M}_{n \alpha' \alpha}({{p}}; j)\, ,
    \label{eqn:matchinginmuaftope}
\end{equation}
where
\begin{align}    
\mathcal{M}_{n \alpha' \alpha}({{p}}; j) \equiv
\langle \psi^{-}_{{{p}} j \alpha'}|\mathcal{O}_n^{(\slashed{\pi})}|\psi^{+}_{{{p}} j \alpha}\rangle\, .
\label{eqn:localmtxele}
\end{align}
For any $k \gg p$ of interest, we compare both sides of Eq.~\eqref{eqn:matchinginmuaftope} and obtain $\widetilde{W}_{n}(k)$. 
The comparison is valid if ${{p}}$ of the scattering states is chosen to be in the validity range of Pionless EFT. 
This kind of procedure is referred to as ``matching,'' often applied to two EFTs of QCD in a shared kinematic region where perturbation theory is justified~\cite{Bauer:2000yr, Eichten:1989zv}. 
For a fixed value of $k$, $\mathcal{M}_{n \alpha' \alpha}({{p}}; j)$ furnish a basis of independent functions of ${{p}}$. 
If more than one Wilson coefficient is needed, one can perform matching at multiple kinematic points of $p$, in order to retrieve more information from the underlying theory.
Once the Wilson coefficients $\widetilde{W}_{n}(k)$ are determined, we can use them in Eq.~\eqref{eqn:OPERhok2} to evaluate $\rho(k)$ for large $k$. By then only long-range structure of nucleus $A$, manifested by $\langle A |\mathcal{O}_n^{(\slashed{\pi})}|A \rangle$, needs to be calculated with Pionless EFT potentials.

\section{Pionless EFT\label{sec:PLEFT}}

We explain in this section the Pionless-EFT side of the story. The Lagrangian terms of Pionless EFT and calculations of few-body systems are well documented in the literature (see, e.g., Ref.~\cite{Hammer:2019poc} for a recent review). 
In Pionless EFT the pions are integrated out, so $NN$ interactions are all contact, which makes analytic evaluations possible for the two-body system.
The first few Pionless EFT Lagrangian terms~\cite{Fleming:1999ee} are
\begin{equation}
    \begin{split}
    \mathcal{L}_\slashed{\pi} =& N^\dagger \left(i\partial_0 + \frac{\overrightarrow{\nabla}^2}{m_N}\right) N \\
    &+ \sum_{\alpha_s} \left[- C_0(\alpha_s)\right] \mathcal{O}_0(\alpha_s)
    + \frac{C_2(\alpha_s)}{8}\mathcal{O}_2(\alpha_s)\\
    &-C_2(SD)\mathcal{O}_2(SD)
+\sum_{\alpha_p} C_2(\alpha_p)\mathcal{O}_2(\alpha_p)+\cdots\label{eqn:PLessLag}
    \end{split}
\end{equation}
where 
$\alpha_s = \cs{1}{0}$ or $\cs{3}{1}$, $\alpha_p = \cp{1}{1}$ or $\cp{3}{0}$ and $\slashed{\pi}$ is dropped when there is no confusion.
Here the $S$-wave contact-interaction operators 
are given by
\begin{align}
\mathcal{O}_{0}(\alpha_s)=&\left[N^{T}P_{i}(\alpha_s)N\right]^{\dagger}\left[N^{T}P_{i}(\alpha_s)N\right]\, ,\\
\mathcal{O}_{2}(\alpha_s)=&\left[N^{T}P_{i}(\alpha_s)N\right]^{\dagger}\left[N^{T}P_{i}(\alpha_s)\overleftrightarrow{\nabla}^{2}N\right]+\text{H.c.}\, , 
\end{align}
where $\overleftrightarrow{\nabla} \equiv \overrightarrow{\nabla}-\overleftarrow{\nabla}$ and the spin-isospin projection matrices are defined as follows:
\begin{align}
P_{i}(\cs{1}{0})&=\frac{\left(i\sigma_{2}\right)\left(i\tau_{2}\tau_{i}\right)}{2\sqrt{2}}\, ,\\
P_{i}(\cs{3}{1})&=\frac{\left(i\sigma_{2}\sigma_{i}\right)\left(i\tau_{2}\right)}{2\sqrt{2}}\, .
\end{align}
The $SD$ mixing contact operator is
\begin{align}
\mathcal{O}_{2}(SD)=&\left[N^{T}P_{i}\left(\cs{3}{1}\right)N\right]^{\dagger}
\left[N^{T}P_{i}\left(\cd{3}{1}\right)N\right]
+\text{H.c.}\, ,
\end{align}
with the $\cd{3}{1}$ projection matrix defined by
\begin{align}
    P_{i}\left(\cd{3}{1}\right)
    &=\frac{3}{4\sqrt{2}}\left(\overleftrightarrow{\nabla}_{i}\overleftrightarrow{\nabla}_{j}-\frac{\delta_{ij}}{3}\overleftrightarrow{\nabla}^{2}\right)P_{j}\left(\cs{3}{1}\right)\, .
\end{align}
The $P$-wave operators are 
\begin{align}
    \mathcal{O}_{2}(\alpha_p)=&\left[N^{T}P_i(\alpha_p)N\right]^{\dagger}\left[N^{T}P_i(\alpha_p)N\right]\, ,
\end{align}
where the $P$-wave projection matrices are as follows:
\begin{align}
    P_i(\cp{1}{1})&=\frac{\sqrt{3}(i\sigma_2)(i\tau_2)}{4\sqrt{2}}\overleftrightarrow{\nabla}_i\, ,\\
    P_i(\cp{3}{0})&=\frac{(i\sigma_2\sigma_j)(i\tau_2\tau_i)}{4\sqrt{2}}\overleftrightarrow{\nabla}_j\, .
\end{align}

The local operators that will appear in the right-hand side of Eq.~\eqref{eqn:OPEScheme} are straightforward to enumerate in Pionless EFT; they are made of, again, only products of the nucleon field $N$. 
Because they must have the same quantum numbers as $N^{\dagger}\left(-\frac{1}{2}\vec{r}\right)N\left(+\frac{1}{2}\vec{r}\right)$: spin $0$, isospin $0$, parity even, among others, and these quantum numbers are precisely those of the Lagrangian terms shown in Eq.~\eqref{eqn:PLessLag}, we can use the Lagrangian operators in the OPE. 
We now list the local operators according to their mass dimensions in ascending order. 
The lowest are one-body operators:
\begin{align}
    \phi_0 &\equiv N^{\dagger}N \, , \; \\
    \phi_2 &\equiv N^{\dagger}\overleftrightarrow{\nabla}^2 N + \text{H.c.} \, ,
\end{align}
followed by two-body operators:
\begin{align}
    &\mathcal{O}_0(\cs{3}{1})\, ,\; \mathcal{O}_0(\cs{1}{0})\, ,  \\
    &\mathcal{O}_2(\cs{3}{1})\, ,\;
    \mathcal{O}_2(\cs{1}{0})\, , \\
    &\mathcal{O}_2(SD)\, , \mathcal{O}_2(\cp{1}{1})\, ,\; \mathcal{O}_2(\cp{3}{0})\, ,\;
    \cdots\, ,\label{eqn:LocOpCP}    
\end{align}
Operators corresponding to interactions in non-deuteron channels, such as $\mathcal{O}_i(\cs{1}{0})$, $\mathcal{O}_i(\cp{3}{0})$, and $\mathcal{O}_i(\cp{1}{1})$, do not contribute to the momentum distribution of the deuteron. 

\begin{figure}
    \centering
    \subfigure[]{
    \includegraphics[scale=0.4]{ 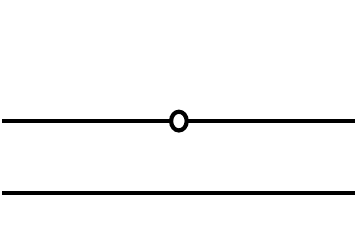}
    }
    \subfigure[]{
    \includegraphics[scale=0.4]{ 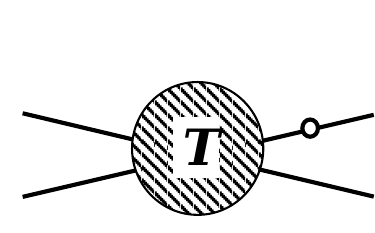}
    }
    \subfigure[]{
    \includegraphics[scale=0.4]{ 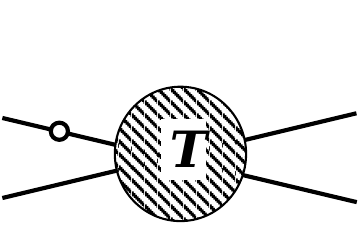}
    }
    \subfigure[]{
    \includegraphics[scale=0.4]{ 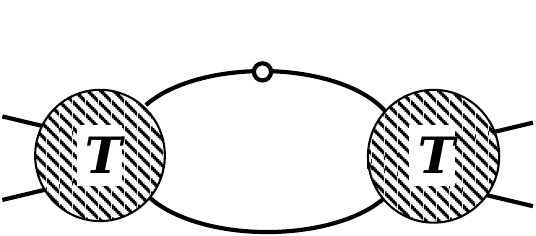}
    }
    \caption{
    Diagrams for the matrix elements of the one-body local operators between scattering states.
    }
    \label{fig:FeynLocalOneBody}
\end{figure}

\begin{figure}
    \centering
    \subfigure[]{
    \includegraphics[scale=0.4]{ 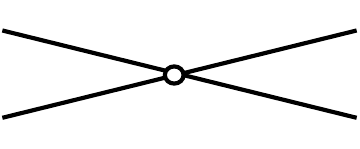}
    }
    \subfigure[]{
    \includegraphics[scale=0.4]{ 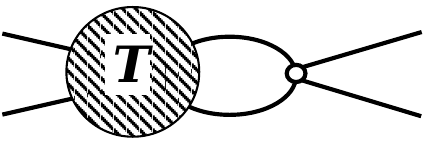}
    }
    \subfigure[]{
    \includegraphics[scale=0.4]{ 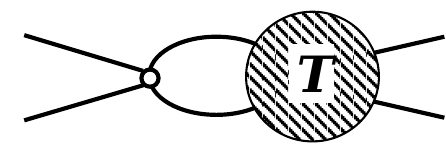}
    }
    \subfigure[]{
    \includegraphics[scale=0.4]{ 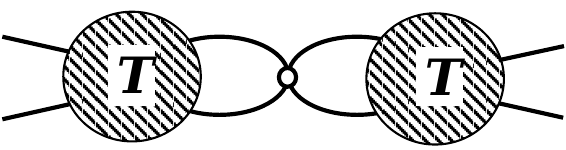}
    }
    \caption{
    Diagram for the matrix elements of two-body local operators in scattering states.}
    \label{fig:FeynLocal2N}
\end{figure}

We proceed to calculate the matrix elements of the local operators between the $\csd$ scattering states $\mathcal{M}_n({{p}})$.
Although matching onto an underlying nuclear theory can be done numerically,  
it is instructive to show the analytic expressions, to the extent we can.
The Feynman diagrams representing the matrix element of the one-body operator $\phi_0$ are shown in Fig.~\ref{fig:FeynLocalOneBody} and those of the two-body operator $\mathcal{O}_0(\cs{3}{1})$ in Fig.~\ref{fig:FeynLocal2N}.
Evaluation for higher-dimension operators is similar. 

One of the elements in evaluating the diagrams of Fig.~\ref{fig:FeynLocal2N} is the four-point function, or the $T$-matrix in $\csd$, represented by the shaded blob. The leading order (LO) $\cs{3}{1}$ potential is simply
\begin{align}
        V^{(0)}(p^{\prime},p)&=C_0^{(0)}\, ,
\end{align}
where we have dropped the channel label $\cs{3}{1}$ for clarity of notations. A separable super Gaussian function is used to regularize the UV part of the $NN$ potentials: 
\begin{equation}
    V^{\Lambda}(p^{\prime},p) = f_R\left(\frac{p^2}{\Lambda^2}\right) V(p^{\prime},p)f_R\left(\frac{{p^{\prime}}^2}{\Lambda^2}\right)\, ,
    \label{eqn:regVpionless}
\end{equation}
where
\begin{equation}
    f_R(x)=e^{-x^2}\, .\label{eqn:gaureg}
\end{equation}
One calculates the LO on-shell $T$-matrix through the LS equation~\eqref{eqn:InHomoLS}, which is much simplified by the fact that $NN$ EFT potentials and the regulator are both separable:
\begin{align}
        T_{\rm LO}({{p}})&=-\frac{4\pi}{m}\frac{1}{-\frac{4\pi }{mC_0}-4\pi\theta_1\Lambda-i{{p}} + O\left(
        \frac{{{p}}^2}{\Lambda}
        \right)}\, ,
        \label{eqn:PLessTLO}
\end{align}
where $\theta_n$ depends on the regularization scheme:
\begin{align}
    \frac{1}{2\pi^2}\int_0^{\infty} dq q^{n-1}f_R^2\left(\frac{q^2}{\Lambda^2}\right)=&\theta_n\Lambda^n\, .
\end{align}
Unless noted otherwise we will not keep track of all the residual cutoff dependence $p/\Lambda$ that will vanish as $\Lambda \to \infty$.

At next-to-leading order (NLO), the $T$-matrix is corrected perturbatively by the following NLO EFT potential:
\begin{align}
    V^{(1)}(p^{\prime},p)&=C_0^{(1)}+C_2^{(0)}(p^2+{p^{\prime}}^2)\, , \,
    \label{eqn:PLessVNLO}
\end{align}
where we have expanded the $C_0$ at each EFT order:
\begin{equation}
    C_0=C_0^{(0)}+C_0^{(1)}+\cdots\, ,
\end{equation}
which we note does not introduce more free parameters to the EFT~\cite{Kaplan:1996xu, Long:2007vp}. 
The NLO $T$-matrix is
\begin{align}
    T_{\rm NLO}({{p}})&
    =\eta \frac{m}{\Lambda}{{p}}^2\, T^2_{\rm LO}(p)
    + C^{(1)}_0\frac{T^2_{\rm LO}(p)}{C_0^2} \nonumber \\
    &\quad + 2C^{(0)}_2\left[\frac{{{p}}^2T^2_{\rm LO}(p)}{{C_0^{(0)}}^2}
    -m\theta_3\Lambda^3\frac{T^2_{\rm LO}(p)}{C_0^{(0)}}\right]\, ,
    \label{eqn:PLessTNLO}
\end{align}
where 
$\eta$ is defined in the expansion of the following integral in ${{p}}/\Lambda$:
\begin{align}
    &\frac{1}{2\pi^2}\int_0^{\infty} dq\frac{mE}{E-\frac{q^2}{m}+i\epsilon}f_R^2\left(\frac{q^2}{\Lambda^2}\right) \nonumber \\
    =&-i\frac{m{{p}}}{4\pi} + \eta \frac{m {{p}}^2}{\Lambda}+O\left(\frac{1}{\Lambda^2}\right).
\end{align}

On the other hand, the LO and NLO $T$-matrix can be written in the effective range expansion as:
\begin{align}
    T_{\rm LO}({{p}})&=-\frac{4\pi}{m}\frac{1}{-\frac{1}{a}-i{{p}}}\, , \label{eqn:ERETLO}\\    
    T_{\rm NLO}({{p}})&= -\frac{4\pi}{m} \frac{-\frac{r_0}{2}{{p}}^2}{\left(-\frac{1}{a}-i{{p}}\right)^2
    }\,    
    ,\label{eqn:ERETNLO}
\end{align}
where $a$ is the scattering length and $r_0$ the effective range in $\cs{3}{1}$ and we have used the Pionless EFT counting $a {{p}} = O(1)$ and $r/a \ll 1$. 
The EFT amplitudes can be renormalized by equating Eq.~\eqref{eqn:PLessTLO} with Eq.~\eqref{eqn:ERETLO}, and Eq.~\eqref{eqn:PLessTNLO} with Eq.~\eqref{eqn:ERETNLO}, respectively. 
This renormalization prescription dictates the following dependence of $C_0$ and $C_2$ on the ultraviolet cutoff $\Lambda$:
\begin{align}
    C_0^{(0)}&=\frac{4\pi a}{m-4\pi m\theta_1\Lambda a}\, ,\label{eqn:pionlessParaC0}\\
    C^{(0)}_2&=\frac{m{C_0^{(0)}}^2}{16\pi}\left(r_0-\frac{8\pi\eta}{\Lambda}\right)\, ,\label{eqn:pionlessParaC2}\\
    C^{(1)}_0&=\frac{\theta_3m^2{C_0^{(0)}}^3\Lambda^3}{8\pi}\left(r_0-\frac{8\pi\eta}{\Lambda}\right)\, .
\end{align}

We start with the matrix element of the one-body operator $N^\dagger N$.
At LO, it is given by 
\begin{equation}
    \begin{split}
        \mathcal{M}_{\phi_0}^{\rm LO}({{p}};\cs{3}{1})
    &\equiv \langle\psi_{{{p}}}^{-};\cs{3}{1}|N^\dagger N|\psi_{{{p}}}^{+};\cs{3}{1}\rangle_{\text{LO}}  \\
    &=(2\pi)^3 \delta^{(3)}(0)+2\frac{T_{\text{LO}}({{p}})}{E-\frac{{{p}}^2}{m}}\\
    &\quad +\frac{im^2}{8\pi}\frac{T^2_{\text{LO}}({{p}})}{p}\, .
\end{split}\label{eqn:M1N}
\end{equation}
To find the Wilson coefficient of $N^\dagger N$, denoted by $\widetilde{W}_{\phi_0}(k)$, we need to identify the $k$-dependent factor that, when multiplied by Eq.~\eqref{eqn:M1N}, can reproduce the exact results of Eqs.~\eqref{eqn:Anonlocal_trivial} and~\eqref{eqn:ANonLocal} for $k \gg p$. The first two terms on the right-hand side of Eq.~\eqref{eqn:M1N} correspond to the disconnected terms in Eq.~\eqref{eqn:Anonlocal_trivial}, so its Wilson coefficient is clearly $0$. The third term $\propto T^2_{\text{LO}}({{p}})/{p}$ can only be related to Eq.~\eqref{eqn:ANonLocal}. However, Eq.~\eqref{eqn:ANonLocal} becomes $\propto T(p, k; E) T(k, p; E)/E^2$ for large $k$ and it will never factorize into the product of $1/p$ and $T^2$ for $p/k \to 0$. Therefore, we conclude that $\widetilde{W}_{\phi_0}(k) = 0$. It means that the one-body local operators do not contribute to the OPE of the momentum distribution for sufficiently large $k$. Similar conclusions are reached in Refs.~\cite{Tropiano:2021qgf,Tropiano:2024bmu}, where a renormalization-group evolution is applied to both the nuclear potential and the nonlocal operator.

The conventional wisdom from the mean-field models of nuclei also suggests that $\widetilde{W}_{\phi_0}(k)$ vanishes for large $k$. 
Had $\widetilde{W}_{\phi_0}(k)$ had a finite value, $\widetilde{W}_{\phi_0}(k) \langle A |N^\dagger N| A \rangle$ could be interpreted as the probability of finding a single nucleon at momentum $k$, with no other nucleon present within distance of order $k^{-1}$.
Since short-range interactions among the nucleons are absent, the mean-field picture applies. 
However, in a mean-field model the nucleon momentum does not go above the Fermi momentum, $k_F \approx 250$ MeV for medium and heavy nuclei~\cite{Moniz:1971mt}.
Therefore, $\rho(k)$ for $k \gg k_F$ must be attributed to few-body interactions at short distance $\sim k^{-1}$.
In the OPE, the short-range correlation of this kind is manifested by two-body local operators and their Wilson coefficients.

The open circles in Fig.~\ref{fig:FeynLocal2N} represent the two-body local operators, corresponding to the following vertex functions:
\begin{align}
        &\mathcal{O}_0(\cs{3}{1})\, :\quad v_0(p^{\prime},p) = 1\, ,\\
        &\mathcal{O}_2(\cs{3}{1})\, : \quad 
        v_2(p^{\prime},p) = -4(p^2+{p^{\prime}}^2)\, .
\end{align}
We choose to regularize the loop integrals involving $v_n$ in the same way as we do the loops involving EFT potentials:
\begin{equation}
    v_n^{\Lambda}(p^{\prime},p)=f_R\left(\frac{p^2}{\Lambda^2}\right) v_n(p^{\prime},p)f_R\left(\frac{{p^{\prime}}^2}{\Lambda^2}\right)\, .
\end{equation}
We have checked that applying different regularization schemes to the local operators induces variations to the Wilson coefficients but does not modify significantly the final result of the nucleon momentum distributions. 
For instance, we have tested regulators such as $f_R(x) = e^{-x}$ and $e^{-x^4}$, and found the discrepancies of the resulting momentum distributions to be order of $10^{-4}$. 

We are now in position to calculate $\mathcal{M}_n(p)$ for the two-body local operators.
At LO, the matrix elements of $\mathcal{O}_0(\cs{3}{1})$ and $\mathcal{O}_2(\cs{3}{1})$ are given by 
\begin{align}
    \nonumber\mathcal{M}_{0}^{\rm LO}({{p}};\cs{3}{1})
    &\equiv \langle\psi_{{{p}}}^{-};\cs{3}{1}|\mathcal{O}_0(\cs{3}{1})|\psi_{{{p}}}^{+};\cs{3}{1}\rangle_{\text{LO}}\\
    &=\frac{{T^2_{\rm LO}({{p}})}}{{C_0^{(0)}}^2}\, ,\label{eqn:localcrlf1LO}\\
    \nonumber\mathcal{M}_{2}^{\rm LO}({{p}};\cs{3}{1})
    &\equiv \langle\psi_{{{p}}}^{-};\cs{3}{1}|\mathcal{O}_2(\cs{3}{1})|\psi_{{{p}}}^{+};\cs{3}{1}\rangle_{\text{LO}}\\
    &= \frac{{T^2_{\rm LO}({{p}})}}{{C_0^{(0)}}^2}
    \left(-8{{{p}}}^2 + 8\theta_3 m\Lambda^3 C_0^{(0)} \right) 
    \, .\label{eqn:localcrlf2LO}    
\end{align}
At NLO, the Pionless NLO potentials are treated perturbatively to construct $|\psi_{{{p}}}^{\pm};\cs{3}{1}\rangle$, or equivalently, they are inserted into the shaded blob 
in Fig.~\ref{fig:FeynLocal2N}. We will need the NLO correction to $\mathcal{M}_{0}^{\rm LO}$:
\begin{align}
    \nonumber\mathcal{M}_{0}^{\rm NLO}({{p}};\cs{3}{1})
    &\equiv\langle\psi_{{{p}}}^{-};\cs{3}{1}|\mathcal{O}_0(\cs{3}{1})|\psi_{{{p}}}^{+};\cs{3}{1}\rangle_\text{NLO}\\
    &=\frac{{T^2_{\rm LO}({{p}})}}{{C_0^{(0)}}^2} \left[-\frac{C^{(1)}_0}{C_0^{(0)}} + 2 \frac{T_{\rm NLO}({{p}})}{T_{\rm LO}({{p}})} \right]
    \, .\label{eqn:localcrlfNLO}
\end{align}
We drop the partial-wave label of $\cs{3}{1}$ in the following discussion to simplify the notations.

$\mathcal{M}_{0}^{\rm LO}({{p}})$, $\mathcal{M}_{2}^{\rm LO}({{p}})$, and $\mathcal{M}_{0}^{\rm NLO}({{p}})$ have dependence on $\Lambda$ which is expected to be absorbed into the Wilson coefficients so that the OPE of $\rho(k)$ is insensitive to the arbitrarily chosen value of $\Lambda$. 
These $\mathcal{M}_{n}({{p}})$ form a basis of independent functions of ${{p}}$, and we would like to reorganize $\mathcal{M}_{n}({{p}})$ so that a new functional basis composed of only ``renormalized'' matrix elements can be formed.
By way of a linear transformation
we choose to construct renormalized $\mathcal{M}_{n}({{p}})$ as follows:
\begin{align}
    \mathcal{M}_{R0}^{\rm LO}({{p}}) &=16\pi^2 \frac{1}{(a^{-1} + i{{p}})^2}\, , \label{eqn:MR0LO}\\
    \mathcal{M}_{R0}^{\rm NLO}({{p}}) &=16\pi^2\frac{\frac{r_0}{2}{{p}}^2}{(a^{-1} + i{{p}})^3}
    \, ,\label{eqn:MR0NLO} \\
    \mathcal{M}_{R2}^{\rm LO}({{p}}) &=16\pi^2\frac{{{p}}^2}{(a^{-1} + i{{p}})^2}\, . \label{eqn:MR2LO}   
\end{align}
The same linear transformation turns the ``bare'' operators $\mathcal{O}_0$ and $\mathcal{O}_2$ into renormalized ones:
\begin{align}
    \mathcal{O}_0^{(R)}  
        &=m^2 C_0^{(0)} \left(C_0^{(0)} + C_0^{(1)}  \right) 
        \mathcal{O}_0\, , \\
    \mathcal{O}_2^{(R)} 
    &=m^2 \left(C_0^{(0)}\right)^2 
    \left(-\frac{1}{8} \mathcal{O}_2
    + \theta_3 m\Lambda^3 C_0^{(0)} 
    \mathcal{O}_0 \right) \, ,    
\end{align}
so that
\begin{align}
\langle\psi_{{{p}}}^{-}|\mathcal{O}_0^{(R)}|\psi_{{{p}}}^{+}\rangle &=\mathcal{M}_{R0}^{\rm LO}({{p}}) +  \mathcal{M}_{R0}^{\rm NLO}({{p}}) + \cdots
     \, , \\
\langle\psi_{{{p}}}^{-}|\mathcal{O}_2^{(R)}|\psi_{{{p}}}^{+}\rangle  
&= \mathcal{M}_{R2}^{\rm LO}({{p}}) + \cdots \, ,
\end{align}
where the ellipses are the EFT corrections in powers of $p/\mhi$.

\begin{figure}
    \centering
    \includegraphics[scale=0.5]{ 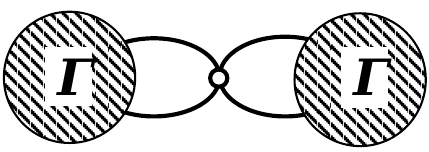} 
    \caption{
    Diagrams for matrix elements of two-body local operators of the deuteron.
    }
    \label{fig:deuMtxlocal}
\end{figure}

The last component from Pionless EFT is the expectation value of local operators $\mathcal{O}_0^{(R)}$ and $\mathcal{O}_2^{(R)}$ in the bound state of the interested nucleus, in this paper, the deuteron. With the computational details relegated to Appendix~\ref{app:DeuMtx}, we only compile the results here. 
The LO three-point vertex function and its NLO correction are given by 
\begin{align}
\Gamma_{\text{LO}}(p) &= \frac{2}{m}\left(\pi a\right)^{-\frac{1}{2}}f_R\left(\frac{p^2}{\Lambda^2}\right)\, , \\
\Gamma_{\text{NLO}}(p) &= \left(\frac{3r_0}{4a}+\frac{C_2^{(0)}}{C_0^{(0)}}p^2\right)\Gamma_{\text{LO}}(p)\, .
\end{align}
The binding energy at each order is obtained along the way:
\begin{align}
    &mB_d^{(0)} = \frac{1}{a^2} \, ,\label{eqn:pionlessBd0}\\
&mB_d^{(1)} =\frac{r_0}{a}\frac{1}{a^2} \label{eqn:pionlessBd1}\, .
\end{align}
With these deuteron properties, one finds the expectation values of $\mathcal{O}_0^{(R)}$ and $\mathcal{O}_2^{(R)}$ in the deuteron, shown diagrammatically in Fig.~\ref{fig:deuMtxlocal}:
\begin{align}
    \langle d|\mathcal{O}_0^{(R)}|d\rangle_{M_J} &=\frac{4}{\pi a}\left(1+\frac{3}{2}\frac{r_0}{a}\right)\, ,\label{eqn:RO0Matx2H}\\
    \langle d|\mathcal{O}_2^{(R)}|d\rangle_{M_J} &=-\frac{4}{\pi a}\frac{1}{a^2} \, .\label{eqn:RO2Matx2H}    
\end{align}
While $\langle d|\mathcal{O}_0^{(R)}|d\rangle$ is computed up to NLO, we stop at LO for $\langle d|\mathcal{O}_2^{(R)}|d\rangle$.

\section{Deuteron in various models\label{sec:models}}

In order to describe the momentum distribution in a kinematic domain that is well beyond the validity range of Pionless EFT, $k \gg m_\pi$, we need inputs from the underlying theory or experimental data. In this section, we assume the underlying theory to be various $NN$ potentials, and show how the OPE and Pionless EFT work together to reproduce the momentum distribution from the underlying nuclear potential. 

The key is the matching procedure described at the end of Sec.~\ref{sec:ope}. The exact matrix elements of the nonlocal operator $\Omega(\vec{k})$ between the in and out $\cs{3}{1}$ states $\mathcal{A}(k,{{p}})$ will be matched with the following OPE terms:
\begin{equation}
\begin{split}
    \mathcal{A}(k, {{p}}) &= {\widetilde{W}^{\text{LO}}_0}(k) \mathcal{M}_{0}^{\rm LO}({{p}}) \\
    &\quad + {\widetilde{W}^{\text{NLO}}_0}(k)\mathcal{M}_{0}^{\rm LO}({{p}}) + {\widetilde{W}^{\text{LO}}_0}(k)\mathcal{M}_{0}^{\rm NLO}({{p}})  \\
    &\quad \quad + {\widetilde{W}^{\mathcal{O}_2}_0}(k)\mathcal{M}_{0}^{\rm LO}({{p}})\\
    &\quad + {\widetilde{W}_2}(k) \mathcal{M}_{2}^{\rm LO}({{p}})
    + \cdots. \label{eqn:SepOPEbare}
\end{split}
\end{equation}
We need to explain ${\widetilde{W}^{\mathcal{O}_2}_0}$.
When $\mathcal{O}_2$ is inserted between low-energy states, some of its effects will appear similar to that of $\mathcal{O}_0$ after $\mathcal{O}_2$ is screened by the $NN$ forces. In other words, $\mathcal{O}_0$ can be renormalized by dressing $\mathcal{O}_2$ with the EFT nuclear interactions. Therefore, ${\widetilde{W}^{\mathcal{O}_2}_0}$ is put in place to account for this ``operator mixing,'' in addition to the EFT expansion of the Wilson coefficients that was already introduced in Eq.~\eqref{eqn:OPERhok2}.

We can choose the bare local operators~[Eqs.\eqref{eqn:localcrlf1LO} - \eqref{eqn:localcrlfNLO}] or the renormalized ones [Eqs.~\eqref{eqn:MR0LO} - \eqref{eqn:MR2LO}] to construct the OPE. In the two models to be studied, one is a toy model suitable for analytic manipulation for which we use the renormalized $\mathcal{O}_n^{(R)}$. The other is the AV18 potential, a realistic nuclear potential frequently used in SRC related studies~\cite{AV18UXVMC14,Yang:2019jwo,Tropiano:2024bmu}, for which we use the bare local operators. 

\subsection{Toy model\label{sec:toy}}

The following separable model describes the $\cs{3}{1}$ $NN$ interaction~\cite{peng22, Beane:2021dab}: 
\begin{equation}
    V_{spr}(p^{\prime},p)=-\frac{4\pi}{m}\frac{g}{\sqrt{p^2+ \gamma^2}\sqrt{{p^{\prime}}^2 + \gamma^2}}\, .\label{eqn:seppot}
\end{equation}
It has the range correction built in but does not have any meson-exchange forces.
Although not as realistic as other nuclear potentials, its simple form allows us to compute analytically the Wilson coefficients.
Due to its separable nature, the on-shell $T$-matrix of the model can be derived with ease:
\begin{equation}
    T_{spr}({{p}})=-\frac{4\pi}{m}\frac{1}{-\gamma+\frac{\gamma^2}{g}+\frac{{{p}}^2}{g}-i{{p}}}\, ,
\end{equation}
from which the effective-range expansion parameters are identified:
\begin{align}
    a^{-1} &= \frac{\gamma}{g}(g - \gamma)\, , \\
    \frac{r_0}{2} &= g^{-1} \, . \label{eqn:rSpr}
\end{align}
In this model, the deuteron only has an $S$-wave component with binding energy given by
\begin{equation}
    B_d = \frac{(g-\gamma)^2}{m} \, .
\end{equation}
As the underlying theory for Pionless EFT, its parameters have the following counting:
\begin{equation}
\begin{split}
    a^{-1} &\sim (g - \gamma) \, ,\\
    \gamma &\sim g \sim \mhi \, . \\
\end{split}\label{eqn:ModelParaScale}
\end{equation}
We will use $1/(a\gamma)$ and $p/\gamma$ to organize the EFT expansion in what follows.

Using Eq.~\eqref{eqn:ANonLocal} we calculate the exact matrix elements of the nonlocal operator $\Omega(\vec{k})$ between the in and out $\cs{3}{1}$ states:
\begin{equation}
    \mathcal{A}_{00}(k,{{p}})
    =\frac{{{p}}^2+\gamma^2}{k^2+\gamma^2}\frac{m^2}{({{p}}^2-k^2)^2}{T^2_{spr}({{p}})}\,.
    \label{eqn:sepnonlocal}
\end{equation}
The momentum distribution in the deuteron $\rho_d(k)$ is given by
\begin{align}
    \rho_d(k) &=\frac{4(g-\gamma)}{\pi m^2}\frac{g^2}{k^2+\gamma^2}\frac{1}{(B_d+\frac{k^2}{m})^2}
    \, .
    \label{eqn:SepRhok2H}
\end{align}

In order to match the exact $\mathcal{A}_{00}(k,{{p}})$ ~\eqref{eqn:sepnonlocal} with the renormalized $\mathcal{M}_{Rn}$, we find that ${\widetilde{W}^{\text{NLO}}_0}$ and ${\widetilde{W}^{\mathcal{O}_2}_0}$ must vanish:
\begin{equation}
    \begin{split}
            \mathcal{A}_{00}(k,{{p}}) &= {\widetilde{W}_0}(k) \left[\mathcal{M}_{R0}^{\rm LO}({{p}}) + \mathcal{M}_{R0}^{\rm NLO}({{p}}) \right]\\
    &\quad 
    + {\widetilde{W}_2}(k) \mathcal{M}_{R2}^{\rm LO}({{p}})
    + \cdots \, ,
    \end{split}
\end{equation}
where 
\begin{align}
       {\widetilde{W}_0}(k) &=\frac{\gamma^2}{k^2+\gamma^2}\frac{1}{k^4} \, , \label{eqn:wil0Model}\\
   {\widetilde{W}_2}(k) &=\left(\frac{2}{k^2}+\frac{1}{\gamma^2}\right){\widetilde{W}_0}(k) \, .
\end{align}
We note that the vanishing ${\widetilde{W}^{\text{NLO}}_0}$ and ${\widetilde{W}^{\mathcal{O}_2}_0}$ are most likely a characteristic of the separable model, and it should not be generally expected of other models. In the small-$k$ limit $k \ll \gamma$, $\widetilde{W}_0(k)$ and $\widetilde{W}_2(k)$ become $\propto k^{-4}$ and $\propto k^{-6}$, respectively, which are in agreement with Refs.~\cite{Braaten:2008uh, Emmons16}.

$\rho_d(k)$ by the OPE/EFT is constructed using Eqs.~\eqref{eqn:RO0Matx2H} and \eqref{eqn:RO2Matx2H}, ${\widetilde{W}_0}(k)$ and ${\widetilde{W}_2}(k)$:
\begin{align}
    \nonumber\rho_d(k)_\text{OPE} =&{\widetilde{W}_0}(k) \frac{4}{\pi a }
    \left[1+\frac{3}{a\gamma} + \mathcal{O}\left(\frac{1}{a^{2}\gamma^{2}}\right)\right]\\ 
    &-{\widetilde{W}_2}(k) \frac{4}{\pi a^3},
     \label{eqn:RhokOPEgamma}
\end{align}
where Eq.~\eqref{eqn:rSpr} has also been used. 
The exact $\rho(k)$ \eqref{eqn:SepRhok2H} can be expanded by the same set of $\widetilde{W}_n(k)$ and the expansion is found to agree with the OPE/EFT result shown in Eq.~\eqref{eqn:RhokOPEgamma}:
\begin{align}
    \rho_d(k)=&{\widetilde{W}_0}(k) \frac{4}{\pi a }\left[1+\frac{3}{a\gamma}+\frac{7}{(a\gamma)^2}+O\left(\frac{1}{a^3\gamma^3}\right)\right] \nonumber\\
    &-{\widetilde{W}_2}(k) \frac{4}{\pi a^3 }\left[1+\frac{5}{a\gamma}+O\left(\frac{1}{a^2\gamma^2}\right)\right] \nonumber \\
    &+\cdots.
    \label{eqn:DifbetOPEandModel}
\end{align}

Although simple, this model illuminates how the OPE/EFT works, especially the roles played by two distinct expansion parameters $1/(ak)$ and $1/(a\mhi)$. For $k$ close to $\mhi$, it is important to incorporate both corrections. For $k \gg \mhi$, ${\widetilde{W}_0}(k)$ may be sufficient and the EFT corrections overweigh inclusion of higher-dimension operators like $\mathcal{O}_2$.

\subsection{
Argonne potential
\label{sec:realistic}}

With a realistic $NN$ potential like AV18, the calculations will be inevitably numerical. We need to explain how the matching is carried out numerically. For a fixed value of $k$ we first use the potential in Eq.~\eqref{eqn:SepOPEbare} to calculate $\mathcal{A}_{00}(k, p_1)$ where $p_1$ is a chosen low-energy kinematic point for matching. Therefore, $p_1 \ll k$ and $p_1$ must be within the domain where Pionless EFT can be applied. If we take into account only $\mathcal{O}_0$ in the OPE and only the LO potential in the EFT expansion, the matching is given by
\begin{align}
\mathcal{A}_{00}(k,p_1) = {\widetilde{W}^{\text{LO}}_0}(k) \mathcal{M}_{0}^{\rm LO}({{p}}_1)\, . \label{eqn:ArgonW0LO}
\end{align}
With the NLO EFT potential kicking in, the matching condition will be adjusted perturbatively~\cite{Long:2007vp, Long:2011xw}. We choose to leave the perfect matching intact at $p_1$ even with the NLO potential, which in turn constrains the NLO Wilson coefficient ${\widetilde{W}^{\text{NLO}}_0}(k)$: 
\begin{align}
{\widetilde{W}^{\text{NLO}}_0}(k)
\mathcal{M}_{0}^{\rm LO}({{p}}_1) + {\widetilde{W}^{\text{LO}}_0}(k)\mathcal{M}_{0}^{\rm NLO}({{p}}_1)=0\, . \label{eqn:ArgonW0NLO}
\end{align}
So far there is only one input from the underlying theory---
$\mathcal{A}_{00}(k, p_1)$.

When $\mathcal{O}_2$ is added, more information is needed from the AV18 potential. In our case, we make use of $\mathcal{A}_{00}$ at another soft momentum $\mathcal{A}(k, p_2)$:
\begin{align}
    &\nonumber \mathcal{A}_{00}(k, p_2) = \left[{\widetilde{W}^{\text{LO}}_0}(k)+{\widetilde{W}^{\text{NLO}}_0}(k)+\widetilde{W}^{\mathcal{O}_2}_0(k)\right]\mathcal{M}_{0}^{\rm LO}({{p}}_2)\\
    &\quad +{\widetilde{W}^{\text{LO}}_0}(k)\mathcal{M}_{0}^{\rm NLO}({{p}}_2)+{\widetilde{W}^{\text{LO}}_2}(k) \mathcal{M}_{2}^{\rm LO}({{p}}_2)\, . \label{eqn:ArgonW2}
\end{align}
In the meantime, it is the job of $\widetilde{W}^{\mathcal{O}_2}_0(k)$ to ensure the matching condition at $p_1$ remains unchanged:  
\begin{align}
    {\widetilde{W}^{\mathcal{O}_2}_0}(k)\mathcal{M}_{0}^{\rm LO}({{p}}_1)+{\widetilde{W}^{\text{LO}}_2}(k) \mathcal{M}_{2}^{\rm LO}({{p}}_1)=0\, .
    \label{eqn:ArgonW0O2}
\end{align}
Solving [Eqs.~\eqref{eqn:ArgonW0LO} - \eqref{eqn:ArgonW0O2}], one obtains four of the Wilson coefficients ${\widetilde{W}^{\text{LO}}_0}(k)$, ${\widetilde{W}^{\text{NLO}}_0}(k)$,  ${\widetilde{W}^{\mathcal{O}_2}_0}(k)$, and ${\widetilde{W}^{\text{LO}}_2}(k)$.
The same procedure is repeated at different values of $k$ if $\rho(k)$ is needed at such $k$'s.

The matching conditions are not unique. They can be modified slightly within the OPE/EFT uncertainty allowed at each order. 
In fact, Refs.~\cite{Gasparyan:2022isg, Peng:2024aiz} discuss a subtle situation where $\mathcal{M}_0(p)$ and $\mathcal{M}_2(p)$ become approximately correlated and renormalization conditions must be treated with additional care, when the EFT interactions are singular long-range attractive potentials, such as the one-pion exchange tensor force. 
Fortunately, Pionless EFT does not suffer from the same problem, at least not in the $NN$ sector.

In the numerical matching performed here we choose $p_1=4$ MeV and $p_2=5$ MeV. The matrix elements of the local operators are also calculated numerically. Although we have demonstrated renormalization of those operators analytically in Sec.~\ref{sec:PLEFT}, there will be residual cutoff dependence in $\mathcal{A}_{00}(p, k)$ and $\rho(k)$. Therefore, we need to specify the cutoff value used in the following Pionless matrix elements: $\Lambda = 800$ MeV.

\begin{figure}
    \centering
    \includegraphics[scale=0.5]{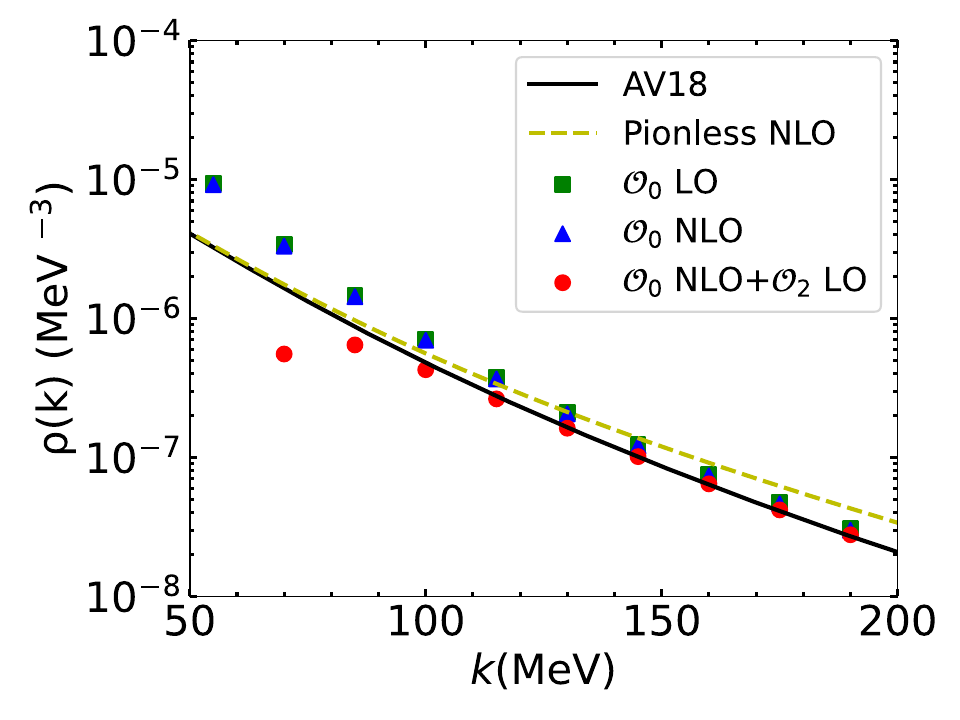} 
    \caption{
    The deuteron single-nucleon momentum distributions calculated with various approaches. The solid line is from the AV18 potential. The dashed line is calculated directly from Pionless EFT. The rest are the results from the OPE-Pionless EFT. See the text for detailed explanation.
    }
    \label{fig:AV18Low}
\end{figure}

Inputs from the scattering states generated by the AV18 potential lead to the predictions of $\rho_d(k)$ by the OPE/EFT.
In Fig.~\ref{fig:AV18Low}, we compare $\rho_d(k)$ of the OPE/EFT with that of the AV18 potential. 
``$\mathcal{O}_0$ LO'' refers to the OPE having only $\mathcal{O}_0$ and the LO EFT interaction. 
For ``$\mathcal{O}_0$ NLO'' the OPE again has only $\mathcal{O}_0$ but the NLO EFT interaction is considered. 
Finally, we add $\mathcal{O}_2$ with the LO interaction to plot ``$\mathcal{O}_0$ NLO + $\mathcal{O}_2$ LO''. 
We also would like to know how Pionless EFT fares without taking any input from the AV18 potential, i.e., a direct calculation by the EFT without assistance of the OPE. 
Represented by the dashed line in Fig.~\ref{fig:AV18Low}, the pure EFT result is shown up to NLO.

\begin{figure}
  \centering
\subfigure[]{\includegraphics[scale=0.5]{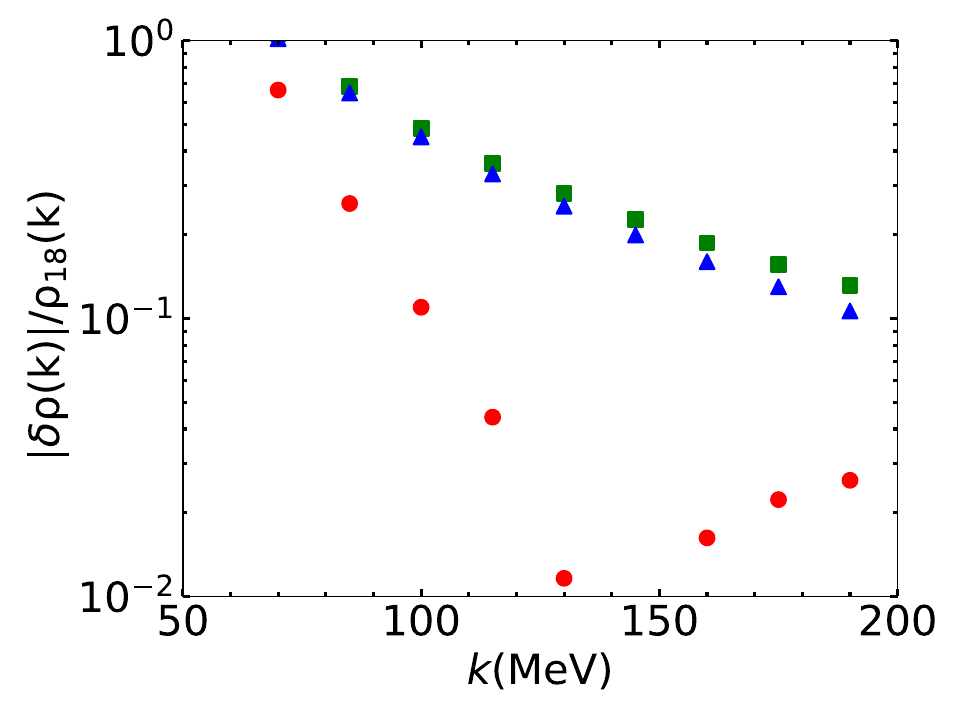}}
\label{fig:ratio50to300}
\subfigure[]{\includegraphics[scale=0.5]{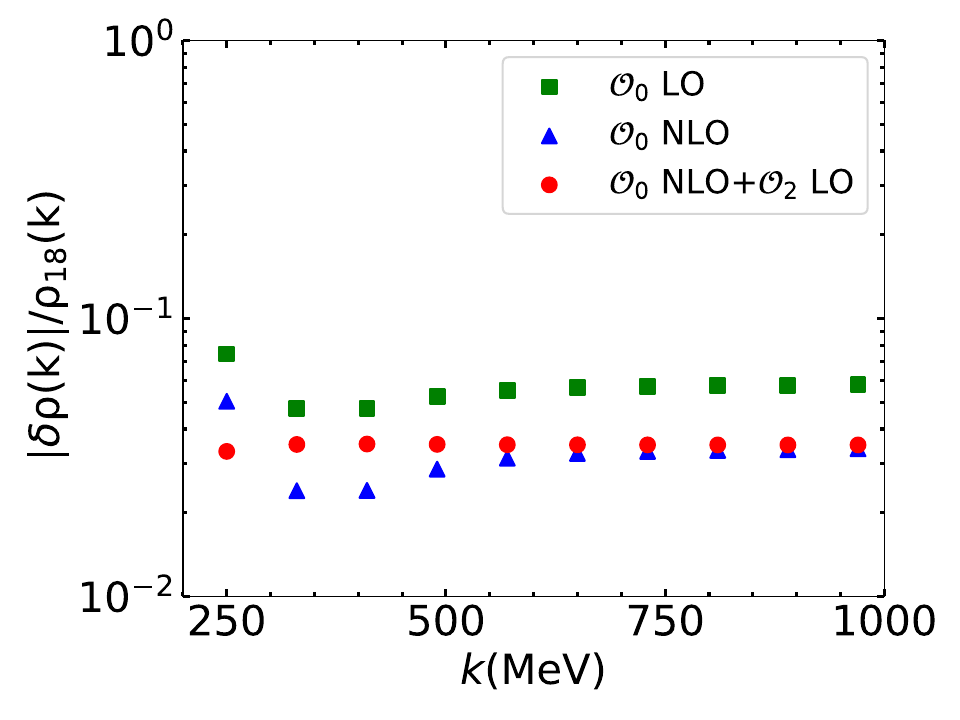}}
\label{fig:ratio300to1000}
\caption{
The relative difference between $\rho_d(k)$ from the OPE/EFT and the AV18 potential. 
See Eq.~\eqref{eqn:RelRhoA} for its definition.}
\label{fig:AV18ErrorRatio}
\end{figure}

It can be more informative to show the relative difference between the OPE/EFT and the model where the absolute scale of $\rho_d(k)$ is too small. 
In Fig.~\ref{fig:AV18ErrorRatio}, we show for each OPE/EFT order
\begin{equation}
\frac{|\delta \rho(k)|}{\rho_\text{18}(k)} \equiv \frac{|\rho_\text{OPE}(k) - \rho_\text{18}(k)|}{\rho_\text{18}(k)}\, , \label{eqn:RelRhoA}
\end{equation}
where $\rho_\text{18}(k)$ is $\rho_d(k)$ evaluated directly with the AV18 potential.

While the OPE works well for large $k$ it breaks down when $k$ decreases to certain values. 
This is clearly demonstrated in Fig.~\ref{fig:AV18Low} in which all the OPE curves diverge away from AV18 for $k \lesssim 100$ MeV. 
But this is no worry because $k$ is so low in this region, the Pionless EFT alone can describe $\rho(k)$.
Toward the opposite limit $k \gg m_\pi$, the expansion in $1/k$ converges fast; therefore, correction due to the NLO EFT interaction proves more important than adding $\mathcal{O}_2$ to the OPE, which is reflected by the proximity of ``$\mathcal{O}_0$ NLO'' to ``$\mathcal{O}_0$ NLO + $\mathcal{O}_2$ LO.'' 
This agrees with the observation made about the toy model where we have also seen that $1/(ak)$ corrections come mainly from the higher-dimension operator $\mathcal{O}_2$. 
In the meantime, it is still worth improving the EFT description because it adds $1/(a\mhi)$ corrections to the Wilson coefficient of $\mathcal{O}_0$ which are not necessarily suppressed for large $k$. In the middle ground $k \sim m_\pi$, corrections from the OPE and EFT interactions both matter.

\section{Discussions and Conclusion\label{sec:con}}

To study a short-range probe of nuclei like SRC physics, we need tools that can separate long-range dynamics governing the internucleon structure and short-range interaction characterizing the probe. The operator product expansion married with Pionless EFT is the tool of our choice in this paper, and the single-nucleon momentum distribution in the deuteron is investigated as an example of its application.

The focus was to demonstrate how the Wilson coefficients of the OPE can be accessed from the $NN$ scattering states prepared by the underlying nuclear model. 
Once they are determined in these two-body calculations by matching the OPE/EFT with the model, applications to many-body bound states will only involve the soft EFT interactions. 

Using the separable toy model for the deuteron, we showed that the momentum distribution by the OPE/EFT is a double expansion in $1/(ak)$ and $1/(a\mhi)$, reproducing term by term $\rho_d(k)$ from the toy model. 
We also tested the OPE/EFT method against the AV18 potential.
The agreement between the OPE/EFT and AV18 was found to be excellent, up to a few percent for $k \gtrsim 250$ MeV.

The successful prediction by the OPE/EFT of the momentum distribution in those deuteron models is based on two principles. 
First, the Wilson coefficients are independent of the external states, so what is extracted from the two-body scattering states can be applied to the many-body bound states. Second, Pionless EFT is able to connect the long-range structure of low-energy scattering states to that of the deuteron because both are within its validity range.

If we ignore the corrections brought by the NLO EFT potentials, the momentum distribution in nucleus $A$ other than the deuteron is expected to have the following expansion:
\begin{equation}
    \rho_A(k) = \widetilde{W}_0(k) \langle A |\mathcal{O}_0| A \rangle_\text{LO} + \widetilde{W}_2(k) \langle A |\mathcal{O}_2| A \rangle_\text{LO} + \cdots.
\end{equation}
When $k$ is large enough, the leading OPE term will be rather accurate and the one-term factorization works really well. 
Therefore, the ratio of $\rho_A(k)$ to  $\rho_d(k)$ is expected to be approximately constant.
But we have seen with both the toy model and AV18 that there exists a medium range of $k$ where $\mathcal{O}_2$ becomes important.
As a result, $\rho_A(k)/\rho_d(k)$ being constant can no longer be taken as a good approximation in those kinematic domains, and its fluctuation will be accounted for by including higher-dimension operators in the OPE.

\acknowledgments

We thank Yu Jia and Yingsheng Huang for useful discussions at the early stage of this work. 
This work was supported in part by the National Natural Science Foundation of China (NSFC) under the Grants No. 12275185 and No. 12335002.

\appendix

\begin{widetext}
    
\section{Rotational property of \texorpdfstring{$\Omega(\vec{k}\,)$}{Omegak}\label{app:RotOmega}}

Suppose that $\bm{R}$ is an infinitesimal rotation around unit axis $\hat{\bm{n}}$. Its effect on a three-momentum is denoted as
\begin{equation}
\vec{k}^{\prime} = \bm{R} \vec{k} \, .
\end{equation}    
We can write its corresponding operator as 
\begin{equation}
    u_{\bm{R}}=1-i\bm{J}_{\hat{\bm{n}}}\phi+O(\phi^2)\, ,
    \label{eqn:rotoperator}
\end{equation}
where $\bm{J}_{\hat{\bm{n}}}$ is the projection of the angular momentum onto $\hat{\bm{n}}$ and $\phi$ is the infinitesimal rotation angle. 
Under the rotation $\bm{R}$ the momentum distribution $\rho(\vec{k}\,)$ is transformed as
\begin{align}
\rho(\vec{k}^{\prime})=&\frac{1}{2J+1} \sum_{M_J}\langle A;J M_J, TM_T |u_{\bm{R}}\Omega(\vec{k})u_{\bm{R}}^{\dagger}|A;J M_J, TM_T\rangle \,
, \nonumber  
\\
=&\rho(\vec{k})-i\phi\frac{1}{2J+1} \sum_{M_J}\langle A;J M_J, TM_T |[\bm{J}_{\hat{\bm{n}}},\Omega(\vec{k})]|A;J M_J, TM_T\rangle\, . \label{eqn:rotrhoEXPAN}
\end{align}
We wish to show that the second term $\delta\rho(\vec{k}\,) \equiv \rho(\vec{k}\,') - \rho(\vec{k}\,)$ in the last line actually vanishes. $\phi J_{\hat{\bm{n}}}$ can be reparametrized as
\begin{equation}
    \phi \bm{J}_{\hat{\bm{n}}}=\alpha \bm{J}_+ +\beta \bm{J}_- +\gamma \bm{J}_z \, ,
\end{equation}
where $\bm{J}_+$ and $\bm{J}_-$ are the raising and lowering operators, and $\bm{J}_z$ is the $z$ component of the angular momentum.
$\delta \rho(\vec{k}\,)$ is then expanded as
\begin{align}
    \delta \rho(\vec{k}\,) \propto\; &\phi \langle JM_J| [\bm{J}_{\hat{\bm{n}}},\Omega(\vec{k})]|JM_J\rangle \nonumber \\
    = &\alpha\sqrt{(J-M_J)(J+M_J+1)} \left(\langle JM_J+1|\Omega(\vec{k})|JM_J\rangle-\langle JM_J|\Omega(\vec{k})|JM_J+1\rangle \right) \nonumber \\
    &+\beta\sqrt{(J+M_J)(J-M_J+1)} \left(\langle JM_J-1|\Omega(\vec{k})|JM_J\rangle-\langle JM_J|\Omega(\vec{k})|JM_J-1\rangle \right)\, .    
\end{align}    
Using the definition of $\Omega(\vec{k})$ \eqref{eqn:omegak}, one can show that $\Omega(\vec{k})$ is Hermitian. It immediately follows that the terms inside the above parentheses are zero, therefore, $\delta \rho(\vec{k}\,)$ vanishes. That is, $\rho(\vec{k}\,)$ depends only on the magnitude of $\vec{k}$.

\section{Deuteron vertex functions\label{app:DeuMtx}}

At Pionless LO, we solve the following homogeneous LS equation for $\Gamma_{\text{LO}}$:
\begin{equation}
    \Gamma_{\text{LO}}(k)=\frac{C_0^{(0)}}{2\pi^2} f_R\left(\frac{k^2}{\Lambda^2}\right) \int dq q^2\frac{f_R\left(\frac{q^2}{\Lambda^2}\right)}{-B_d^{(0)}-\frac{q^2}{m}}\Gamma_{\text{LO}}(q)\, ,\label{eqn:GammaLOapp}
\end{equation}
which clearly indicates $\Gamma_{\text{LO}}(k)\propto f_R\left(\frac{k^2}{\Lambda^2}\right)$. Using this on both sides of the homogeneous LS equation, one can obtain the LO binding energy:
\begin{equation}
    mB_d^{(0)}=\frac{1}{a^2}\, .
\end{equation}
Next, we use the normalization convention equation~\eqref{eqn:NormRhok} to find
\begin{equation}
    \Gamma_{\text{LO}}(k)
    =\frac{2}{m}\frac{f_R\left(\frac{k^2}{\Lambda^2}\right)}{\sqrt{\pi a}}\, .
\end{equation}

The NLO correction, when treated perturbatively, to the LS equation reads
\begin{equation}
    \begin{split}
    \Gamma_{\text{NLO}}(k)=\frac{f_R\left(\frac{k^2}{\Lambda^2}\right)}{2\pi^2}
    \int dq q^2 
    &\Bigg[ \frac{C_0^{(0)}}{-B_d^{(0)}-\frac{q^2}{m}} \Gamma_{\text{NLO}}(q)
    + \frac{C_0^{(1)}+C_2^{(0)}(k^2+q^2)}{-B_d^{(0)}-\frac{q^2}{m}} \Gamma_{\text{LO}}(q)\\
    &\qquad + \frac{C_0^{(0)}B_d^{(1)}}{\left(-B_d^{(0)}-\frac{q^2}{m}\right)^2} \Gamma_{\text{LO}}(q) \Bigg] 
    f_R\left(\frac{q^2}{\Lambda^2}\right) \,
        .\label{eqn:GammaNLOapp_pert}
    \end{split}
\end{equation}
We parametrize $\Gamma_{\text{NLO}}(k)$ as
\begin{equation}
    \Gamma_{\text{NLO}}(k)=\left[\Gamma_{\text{NLO}}(0)+\mathcal{Y}k^2\right]f_R\left(\frac{k^2}{\Lambda^2}\right)\, , \label{eqn:GammaNLOtemp}
\end{equation}
where $\mathcal{Y}$ can be evaluated:
\begin{equation}
\begin{split}
    \mathcal{Y}=&\frac{1}{2\pi^2}\int dq q^2\frac{C_2^{(0)}}{-B_d^{(0)}-\frac{q^2}{m}}f_R\left(\frac{q^2}{\Lambda^2}\right)\Gamma_{\text{LO}}(q)\\
    =&\frac{C_2^{(0)}}{C_0^{(0)}}\Gamma_{\text{LO}}(0)\, .
\end{split}
\end{equation}
Using Eq.~\eqref{eqn:GammaNLOtemp} in Eq.~\eqref{eqn:GammaNLOapp_pert}, one obtains the NLO correction to the binding energy:
\begin{equation}
    mB_d^{(1)}=\frac{r_0}{a}\frac{1}{a^2}\, .
\end{equation}
To have the normalization condition satisfied perturbatively at NLO, we must have the following identity:
\begin{equation}
   2 \int dk k^2\frac{\Gamma^2_{\text{LO}}(k)}{(B_d^{(0)} + \frac{k^2}{m})^2}
   \left[ \frac{\Gamma_{\text{NLO}}(k)}{\Gamma_{\text{LO}}(k)} - \frac{B_d^{(1)}}{B_d^{(0)}+\frac{k^2}{m}} \right] = 0 \, ,
\end{equation}
which at last fixes $\Gamma_{\text{NLO}}(k)$:
\begin{equation}
    \Gamma_{\text{NLO}}(k) = \left(\frac{3r_0}{4a}+\frac{C_2^{(0)}}{C_0^{(0)}}k^2\right)\Gamma_{\text{LO}}(k)\, .
\end{equation}

\end{widetext}

\bibliography{eftrun.bib}

\end{document}